\documentclass{emulateapj}

\newcommand{\gray}{$\gamma$-ray}

\newcommand{\Dxx}{D_{xx}}
\newcommand{\hi}{H {\sc i}}

\newcommand{\adv}{Adv.\ Space Res.}

\newcommand{\jcap}{JCAP}
\newcommand{\pubbook}[5]{#5, #1, #2 #3, #4}
\newcommand{\pubjournal}[5]{#4, #1, #2, #3}

\newcommand{\pubproc}[8]{#1 #7 {\it #2} ({\it #4}) #3 (#5) p~#6}

\shorttitle{IC Origin of the Galactic Ridge Emission}
\shortauthors{Porter et al.}

\begin{document}

\title{Inverse Compton Origin of the Hard X-Ray and Soft Gamma-Ray Emission from the Galactic Ridge}

\author{Troy A. Porter}
\affil{
  Santa Cruz Institute for Particle Physics,
  University of California, Santa Cruz, CA 95064
}

\author{Igor V. Moskalenko\altaffilmark{1}}

\affil{
   Hansen Experimental Physics Laboratory, 
   Stanford University, Stanford, CA 94305
}
\altaffiltext{1}{Also Kavli Institute for Particle Astrophysics and Cosmology,
Stanford University, Stanford, CA 94309}

\author{Andrew W. Strong and Elena Orlando}
\affil{
  Max-Planck-Institut f\"ur extraterrestrische Physik,
  Postfach 1312, D-85741 Garching, Germany
}

\and

\author{L. Bouchet}
\affil{CESR-CNRS, 9 Av. du Colonel Roche, 31028 Toulouse Cedex 04, France
}

\begin{abstract}
A recent re-determination of the non-thermal component of the 
hard X-ray to soft \gray{} emission from the 
Galactic ridge, using the SPI instrument on the 
{\it INTErnational Gamma-Ray Astrophysics Laboratory} (INTEGRAL) 
Observatory, is shown to be well reproduced as inverse-Compton emission from 
the interstellar medium.
Both cosmic-ray primary electrons and secondary electrons and positrons 
contribute to the emission. 
The prediction uses the GALPROP model and includes
a new calculation of the interstellar radiation field.
This 
may solve a long-standing mystery of the origin of this emission,
and potentially opens a new window on Galactic cosmic rays.
\end{abstract}

\keywords{
elementary particles ---  
cosmic rays ---
gamma-rays: theory 
}

\section{Introduction}

The Galactic ridge is known to be an intense source of 
continuum hard X- and \gray{} emission. 
The hard X-ray emission was
discovered in 1972 \citep{bleach72}, and interstellar emission has 
subsequently been observed 
by HEAO-1, Tenma (ASTRO-B), 
ASCA, Ginga, RXTE, OSSE 
\citep{Worrall1982,Koyama1986,purcell96,kinzer99,kinzer01}, 
and most recently by Chandra and XMM-Newton.
The \gray{} observations started with the OSO-III satellite in 
1968, followed by SAS-2 in 1972, COS-B (1975--1982) and COMPTEL and EGRET
on the CGRO (1991--2000).  
With COMPTEL and EGRET the improvement in data quality was sufficient 
to allow such studies to be performed in much greater detail.
The Galactic diffuse emission is a major study objective for INTEGRAL
(in orbit since 2002) and the GLAST LAT (to be launched in 2008) 
\citep{Michelson2007,Ritz2007}.
Each of these experiments represents a significant 
leap forward with respect to its predecessor.

Continuum emission of diffuse, interstellar nature is expected in the 
hard X-ray and \gray{} regime from the physical processes of positron 
annihilation 
(through intermediate formation of positronium), 
inverse-Compton (IC)
scattering and bremsstrahlung from cosmic-ray (CR) electrons and positrons, and
via decay of neutral pions produced by interactions of CR nuclei with the 
interstellar gas. 
Positron annihilation in flight (continuum) may
contribute in the few MeV range \citep{Beacom06}.
For the non-positronium continuum, hard X-rays from bremsstrahlung emission 
imply a luminosity in CR electrons which is unacceptably large 
\citep[see e.g.][]{dogiel02a}. 
Composite models have been proposed which incorporate 
thermal and nonthermal components from electrons accelerated in supernovae 
or the ambient interstellar turbulence \citep{valinia00b}.
At MeV energies the origin of the emission is also uncertain \citep{SMR00}.

Alternatively, the origin of the ridge emission could be attributed to
a population of sources too weak to be detected individually, and
hence would not be 
truly interstellar. 
In general, \gray{} telescopes 
have inadequate spatial resolution to clarify this issue.
For X-rays, important progress in this area was made by ASCA
\citep{kaneda97} and Ginga \citep{yamasaki97}.  
More recently,
high-resolution imaging in X-rays (2--10 keV) with Chandra
\citep{ebisawa01,ebisawa05} has claimed to prove the existence of a
truly diffuse component. 
Similiarly, it has been claimed from
an analysis of XMM-Newton data \citep{hands04} that 80\% of the Galactic-ridge
X-ray emission is probably diffuse, and only 9\% can be accounted for
by Galactic sources, the rest being extragalactic in nature.  
However, more
recently \citet{Revnivtsev2006}, with RXTE PCA data, and \citet{Krivonos2007}, 
using INTEGRAL/IBIS and RXTE data, argue convincingly that below 50 keV 
all the ``diffuse''
emission can be accounted for by a Galactic population of sources,
mainly magnetic cataclysmic variables; see also \citet{Revnivtsev2007a} 
and \citet{Revnivtsev2007b}.

At higher energies, an extensive study of the diffuse Galactic 
\gray{} emission in the
context of CR propagation models has been carried out by
\citet{SMR04b,SMR00}. 
This study confirmed that models based 
on locally measured electron and nuclei spectra and synchrotron
constraints are consistent with \gray{} measurements in the 30 MeV --
500 MeV range; outside this range deviations from the data are apparent.
The puzzling excess in the EGRET diffuse emission data above 1 GeV
relative to that expected \citep{SMR00,Hunter97} has shown up in all
models that are tuned to be consistent with the locally measured CR
nuclei and electron spectra \citep{M04rev,SMR04b}. 
The excess has shown up in all directions, not only in the Galactic plane.
This implies that the GeV excess is not a feature
restricted to the Galactic ridge or the gas-related emission. 
A simple
re-scaling of the components ($\pi^0$-decay, IC, bremsstrahlung) does not
improve the fit in any region, 
since the observed peak is at an energy
higher than the $\pi^0$-peak. 
For recent reviews see 
\citet{M04rev}, \citet{Strong2007}, and references therein.

Assuming that the GeV excess is not an instrumental 
artefact\footnote{A discussion of uncertainties and possible sources of 
error associated with determining the diffuse Galactic \gray{} emission 
using EGRET data is available in \citet{moskalenko2007a}}, 
the so-called ``optimised model,'' which explains the GeV excess
in terms of CR intensity variations in the Galaxy, has been proposed by 
\citet{SMR04b}. 
It reproduces the spectrum of the diffuse
\gray{s} in \emph{all directions}, as well as the
latitude and longitude profiles for the whole EGRET energy range 30
MeV -- 50 GeV at the cost of relaxation of the restrictions imposed by
the measurements of local CR proton and electron spectra.
At lower energies, the predictions of this 
model have never
been tested because of the lack of good data.  

The study of the Galactic-ridge continuum X-ray emission is a key goal 
of the INTEGRAL mission. 
The high spectral resolution combined 
with its imaging capabilities promises new 
insights into the nature of this enigmatic radiation.
Previous work based on initial, smaller sets of INTEGRAL/SPI observations 
have reported
the detection of diffuse emission at a level consistent with previous 
experiments 
\citep{strong03a,strong03b}.
However, statistical and systematic errors were large, due in part to the 
uncertainty in the point-source contribution.  
Meanwhile, a new analysis of INTEGRAL/IBIS data 
\citep{lebrun04,terrier04} showed
that, up to 100 keV, indeed a large fraction of the total emission from the 
inner
Galaxy is due to sources. 
\citet{strong04} used the source catalogue from this work 
(containing 91 sources)             
as input to SPI model fitting, giving a more 
solid basis 
for the contribution of point sources in such an analysis.
This exploited the complementarity of the instruments on
INTEGRAL for the first time in the context of diffuse emission.
The SPI analysis by \citet{bouchet05} gave a rather lower 
50--1000 keV power-law continuum
than \citet{strong05}, but the errors were large in the early datasets.
Now, a new analysis \citep{bouchet07} with 3 times as much SPI exposure
gives better statistics, background handling, and point-source subtraction.

In the present paper we focus on energies above
50 keV where sources do not appear to be important because of the
rapid cut-off in the spectra of the majority, and the relatively small
number of hard-spectrum sources.
We use the GALPROP model together with a new model for the 
Galactic interstellar radiation field (ISRF) to solve a long-standing 
mystery of the origin of the hard X-ray emission -- IC emission 
from CR electrons and positrons -- 
and to build a model of the Galactic diffuse emission in the energy range from
keV to TeV energies, thus covering more that 10 orders of magnitude in energy.

Primary CR electrons are directly accelerated in CR sources like supernova
remnants or pulsars.
Secondary electrons and positrons are 
produced via interactions of energetic nuclei with interstellar gas, and 
are usually considered a minor component of CRs. 
This is indeed the case in the heliosphere where the
positron to all-electron ratio is small at all energies,
$e^+/(e^+ + e^-)_{tot} \sim 0.1$. 
However, as we will show, the combined secondary electron/positron 
flux in the
interstellar medium (ISM) is more than half of the
primary CR electron flux at $\sim$1 GeV energies and below.
This leads to a considerable
contribution by secondary positrons and electrons to the diffuse \gray{}
flux via IC scattering and bremsstrahlung and significantly, 
by up to a factor of $\sim$2, 
increases the flux of diffuse Galactic emission below $\sim$100 MeV. 
Secondary positrons and electrons are, therefore, directly seen in hard X-rays
and \gray{s}.

\section{GALPROP code}
\begin{figure*}
\centerline{
\includegraphics[width=3.5in]{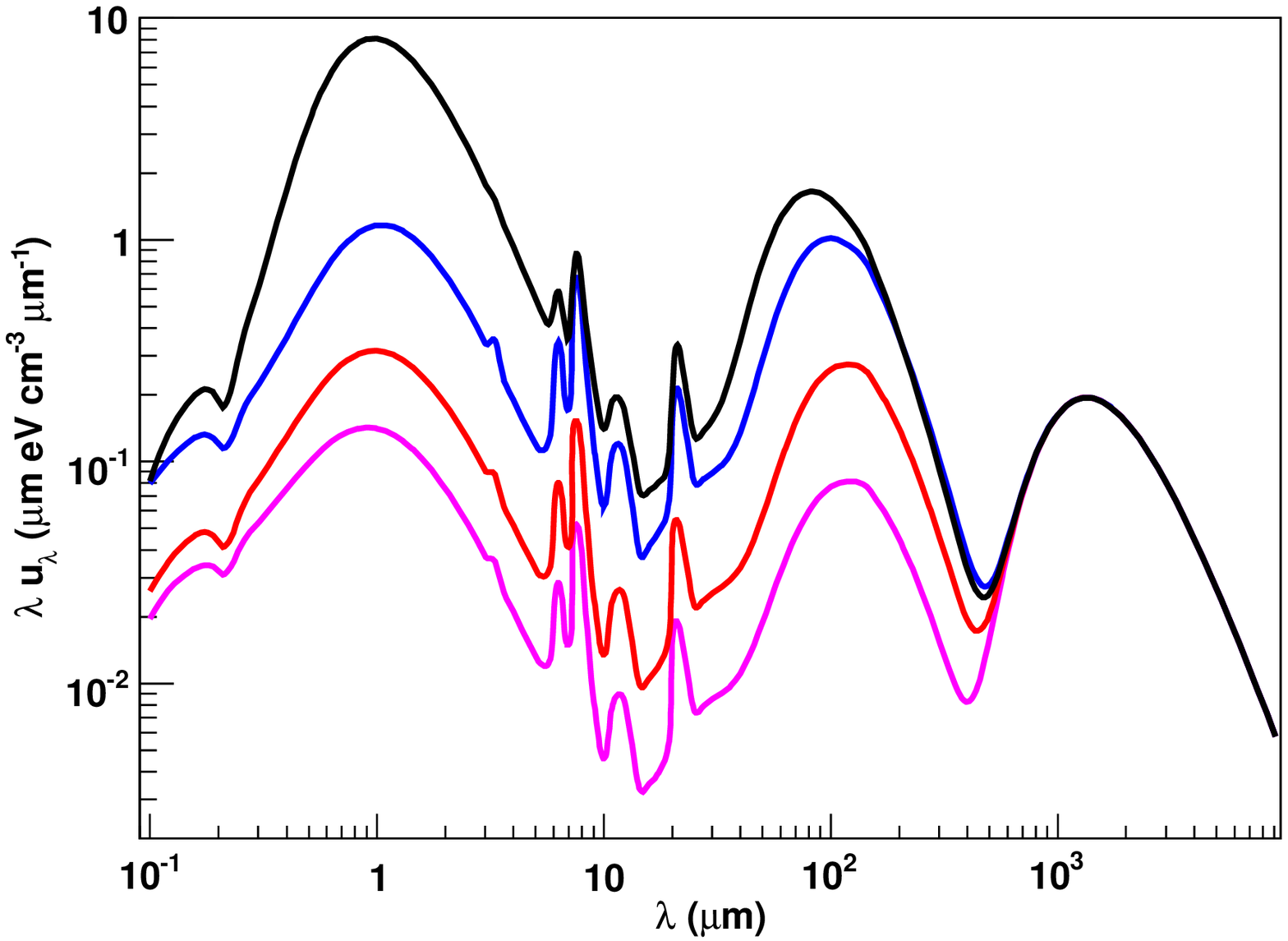}
\includegraphics[width=3.5in]{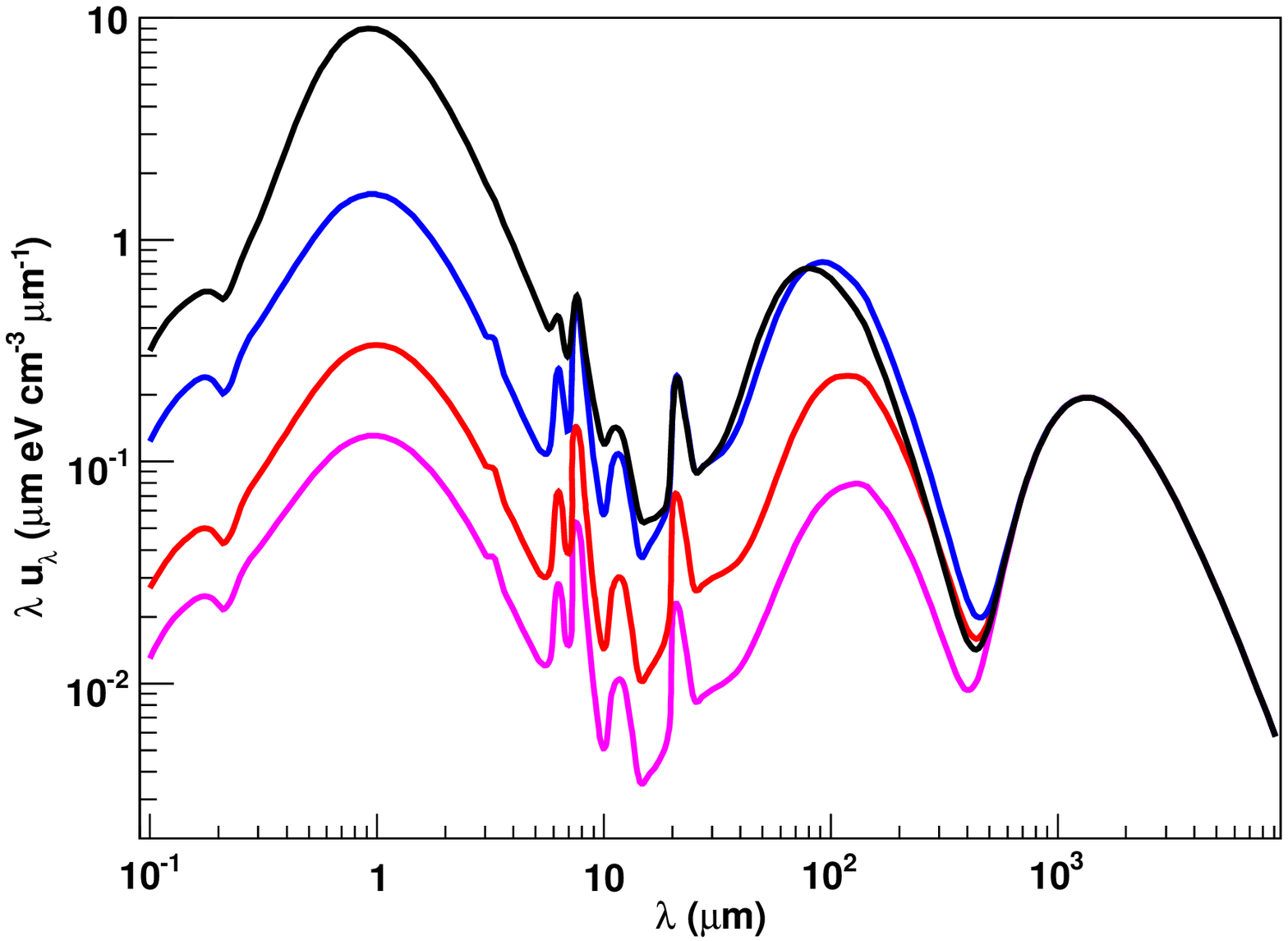}}
\caption{Spectral energy distribution of the MW ISRF in the Galactic plane.
Line colouring: black, $R = 0$ kpc; blue, $R = 4$ kpc; red, $R = 8$ kpc;
magenta, $R = 12$ kpc.
{\it Left:} maximum metallicity gradient; {\it Right:}, minimum metallicity
gradient. The cosmic microwave background (CMB) is included in both figures
and dominates the SED for wavelengths $\lambda \gtrsim 600$ $\mu$m.}
\label{fig:ISRF}
\end{figure*}

The GALPROP code \citep{SM98} was created 
to enable simultaneous predictions of all
relevant observations including CR nuclei, electrons and positrons,
\gray{s} and synchrotron radiation.

We give a very brief summary of GALPROP; for details we refer the
reader to the relevant papers
\citep{SMR00,SMR04b,MS98,MS00,M02,SM98,Ptuskin06} and the dedicated
website\footnote{\tt http://galprop.stanford.edu}. 
The GALPROP code solves the CR transport equation with a given source 
distribution and boundary conditions for all CR species. 
This
includes a galactic wind (convection), diffusive reacceleration in the 
ISM, energy losses, nuclear fragmentation, 
radioactive decay,
and production of secondary particles and isotopes.
The numerical solution of the transport equation is based
on a Crank-Nicholson \citep{Press92} implicit second-order scheme. 
The spatial boundary conditions assume free particle escape. 
Since the grid involves a 3D $(R,z,p)$ or 4D $(x,y,z,p)$ problem
(spatial variables plus momentum)
``operator splitting'' is used to handle the implicit solution.
For a given halo size the diffusion coefficient, as a function of
momentum and the reacceleration or convection parameters, is determined
by the boron-to-carbon ratio data. 
If reacceleration is included, the momentum-space diffusion
coefficient $D_{pp}$ is related to the spatial coefficient $\Dxx$ 
($= \beta D_0\rho^{\delta}$)
\citep{berezinskii90,Seo1994}, where $\delta=1/3$ for a Kolmogorov spectrum
of interstellar turbulence or $\delta=1/2$ for a Kraichnan cascade,
$\rho$ is the magnetic rigidity, $D_0$ is a constant, and $\beta= v/c$.
Production of secondary positrons and electrons is calculated using 
a formalism described in \citet{MS98} with a correction by \citet{Kelner06}.
The \gray{s} are calculated using the propagated CR distributions, 
including a contribution from secondary
particles such as positrons and electrons from inelastic processes in the ISM 
that increases the \gray{} flux at MeV energies \citep{SMR04b}.
Gas-related \gray{} intensities are computed from the emissivities as a
function of $(R,z,E_\gamma)$ using the column densities of \hi\ and
H$_2$ for galactocentric annuli based on 21-cm and CO surveys included
in the GALPROP model. 
Neutral pion production is calculated using the method given by 
\citet{Dermer86a,Dermer86b} as described in \citet{MS98}, or 
using a parameterisation developed by \citet{kamae2005};
bremsstrahlung is calculated
using a formalism by \citet{KochMotz59} as described in \citet{SMR00}.
The IC scattering is treated using the appropriate cross section 
for an
anisotropic radiation field developed by \citet{MS00}
using the full angular distribution of the ISRF.

Cross-sections are based on the extensive LANL database,
nuclear codes, and parameterisations
\citep{Mashnik2004}. 
Starting with the heaviest primary nucleus considered (e.g.\
$^{64}$Ni) the propagation solution is used to compute the source term
for its spallation products, which are then propagated in turn, and so
on down to protons, secondary electrons and positrons, and
antiprotons.  
The inelastically scattered protons and antiprotons are treated 
as separate components (secondary protons, tertiary antiprotons).
In this way secondaries, tertiaries, etc., are included.
(Production of $^{10}$B via the $^{10}$Be-decay channel is  important
and requires a second iteration of this procedure.)  
GALPROP includes
K-capture and electron stripping processes, where a nucleus with an
electron (H-like) is considered a separate species because of the
difference in lifetime, and knock-on electrons.  
Primary
electrons are treated separately.  
Normalisation of protons, alphas, 
and electrons to experimental data is provided (all other isotopes
are determined by the source composition and propagation). 
Gamma rays
are computed using interstellar gas data (for
$\pi^0$-decay and bremsstrahlung) and the ISRF model (for IC).
The synchrotron emission is computed using the Galactic magnetic field model.
Spectra of all species on the chosen grid and the \gray{}
and synchrotron skymaps are output in standard astronomical formats
for comparison with data.  
Recent extensions to GALPROP include
non-linear wave damping \citep{Ptuskin06} 
and a dark matter package to allow for the propagation of 
WIMP annihilation
products and calculation of the corresponding synchrotron and \gray{} skymaps;
an interface between GALPROP and the DarkSUSY code \citep{Gondolo2004} 
will be implemented in the near future to allow direct calls of GALPROP 
from within DarkSUSY.

The optimised model \citep{SMR04b} is used to calculate the 
diffuse emission in the range
10 keV -- TeV energies.
The CR source distribution is based on 
the Galactic pulsar distribution \citep{Lorimer2004}, 
while the $X_{\rm CO}$-factors, 
$X_{\rm CO}=N({\rm H}_2)/W_{\rm CO}$, are variable, increasing towards the 
outer Galaxy, and fully compatible with the expected variations based
on the metallicity gradient and COBE data \citep{SMR04d}.
Such a model reproduces the diffuse Galactic \gray{} emission for the 
whole sky as well as the radial gradient of diffuse Galactic \gray{} emissivity.

\begin{figure*}
\centerline{
\includegraphics[width=3.50in]{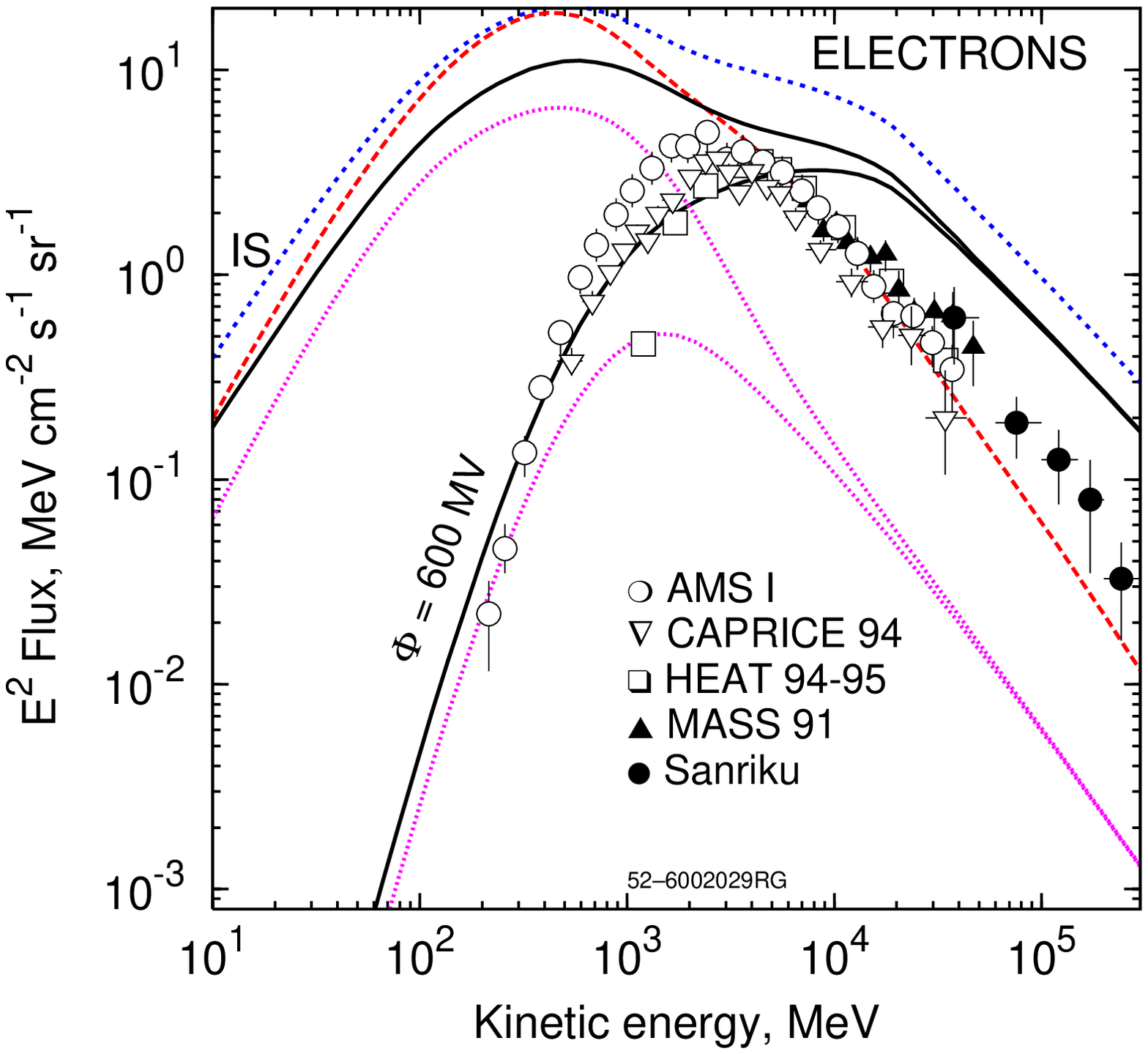}
\includegraphics[width=3.50in]{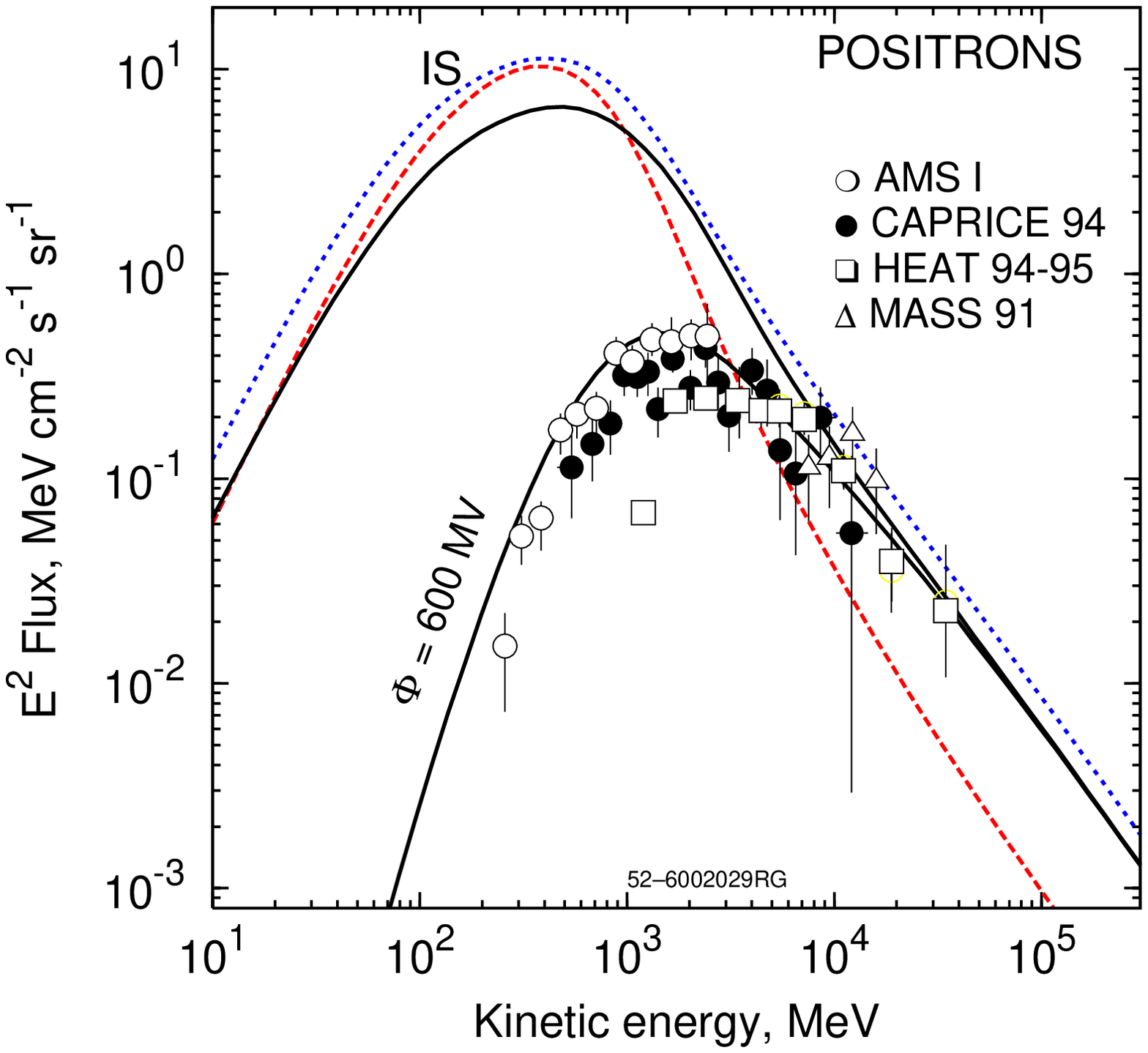}}
\caption{Spectra of CR electrons and positrons in the Galactic plane, 
as predicted by the adopted optimised GALPROP model.
{\it Left:} Total (primary + secondary) and secondary electrons; 
{\it Right:} Secondary positrons.
Interstellar spectra (IS):
$R = 0$ kpc (red long dashes), 
$R = 4$ kpc (blue short dashes), 
$R = 8.5$ kpc (black solid), also shown  modulated to 600 MV.
Secondary electrons are shown separately as magenta lines 
(IS and modulated) on the left panel at $R = 8.5$ kpc.}
\label{fig:electrons_positrons}
\end{figure*}

\section{Interstellar radiation field}


The Galactic ISRF is the result of emission
by stars, and the scattering, absorption, and re-emission of absorbed
starlight by dust in the ISM.
The most detailed calculation to date \citep{SMR00}, which
includes spatial and wavelength dependence over the whole Galaxy,
has been widely used.
The \citet{SMR00} 
model uses emissivities based on stellar populations based on
COBE/DIRBE fits by \citet{Freudenreich1998} and the SKY model of 
\citet{Wainscoat1992} together with COBE/DIRBE derived 
infrared emissivities \citep{Sodroski1997,Dwek1997}.
Subsequent to this work new relevant astronomical
information on stellar populations, Galactic structure, and interstellar dust
has become available, motivating a re-evaluation of the ISRF. 
We briefly describe our calculation of the ISRF; 
further details can be found in \citet{MPS2006} and \citet{PMS2006}. 

The fundamental factors influencing the ISRF 
are the luminosity distribution from the stellar populations of the Galaxy and
the radiative transport of the star light through the ISM.
The interstellar dust absorbs and scatters the star light in the ultraviolet 
(UV) and optical, and re-emits the absorbed radiation in the infrared.

In our model, we represent the stellar distribution by four 
spatial components: the thin and thick disc, the bulge, and the 
spheroidal halo. 
We follow \citet{Garwood1987} and \citet{Wainscoat1992}
and use a table of stellar spectral types comprising
normal stars and exotics to represent the luminosity 
function (LF) for each of the spatial components.
The spectral templates for each stellar type are taken from the 
semi-empirical library of \citet{Pickles1998}.
The normalisations per stellar type are obtained by adjusting the 
space densities to reproduce the observed LFs in the V- and K-band for
the thin disc.
The LFs for the other spatial components are obtained by adjusting 
weights per component for each of the stellar types relative to the 
normalisations obtained for the thin disc LF.

We assume a dust model including graphite, polycyclic aromatic hydrocarbons
(PAHs), and silicate.
Dust grains in the model are spherical and the absorption and scattering 
efficiencies for graphite, PAHs, and silicate grains are taken from 
\citet{Li2001}.
The dust grain abundance and size distribution are taken from 
\citet{Weingartner2001} (their best fit Galactic model).
We assume a purely neutral ISM.
We consider only coherent scattering, and a Henyey-Greenstein angular
distribution function \citep{Henyey1941} is used in the scattering calculation.
The stochastic heating of grains smaller than $\sim$0.1 $\mu$m
is treated
using the ``thermal continuous'' approach of \cite{Draine2001};
we calculate the equilibrium heating of larger dust grains by balancing
absorption with re-emission as described by \cite{Li2001}.

Dust follows the Galactic gas distribution and we assume uniform 
mixing between the two in the ISM \citep{Bohlin1978}.
The dust-to-gas ratio scales with the Galactic metallicity gradient.
Estimates for the Galactic [O/H] gradient vary in the range $0.04-0.07$
dex kpc$^{-1}$ \citep[][and references therein]{SMR04d}.
The variation of the metallicity gradient influences
the redistribution of the mainly UV and blue 
component of the ISRF into
the infrared: increased metallicity implies more dust, which enhances the 
absorption of the star light.
The variation in the infrared component affects the 
emission in the hard X-rays (see below).
Therefore, we consider two ISRFs
corresponding to a maximal case 
of $0.07$ dex kpc$^{-1}$ and a minimal case with no gradient.

The ISRF is calculated for a cylindrical geometry with azimuthal symmetry.
The maximum radial extent is $R_{\rm max} = 20$ kpc with the maximum height
above the galactic plane $z_{\rm max} = 5$ kpc.
The radiative transport is performed using the so-called partial 
intensity method \citep{Kylafis1987,Baes2001}.

Figure~\ref{fig:ISRF} shows the spectral energy distributions (SEDs) in the 
Galactic plane for selected galactocentric radii for the maximal 
metallicity gradient (left) and no metallicity gradient (right).
An increased metallicity gradient reduces the UV in the inner Galaxy 
significantly -- by up to a factor of 3 for $\lambda \lesssim 0.3$ $\mu$m -- 
which is 
redistributed into the infrared.
The infrared emission for the inner Galaxy for the maximum metallicity 
gradient is a factor of $\sim 2$ higher than for the case of no metallicity
gradient.
For the outer Galaxy the ISRFs calculated for the two cases differ less 
dramatically.
For the case of the maximal metallicity gradient the UV emission is higher 
than the no gradient case because there is less dust in the outer 
Galaxy.
In turn, this results in less emission in the infrared than for the maximal 
gradient case.

\section{Results}

\begin{figure*}
\centerline{
\includegraphics[width=3.5in]{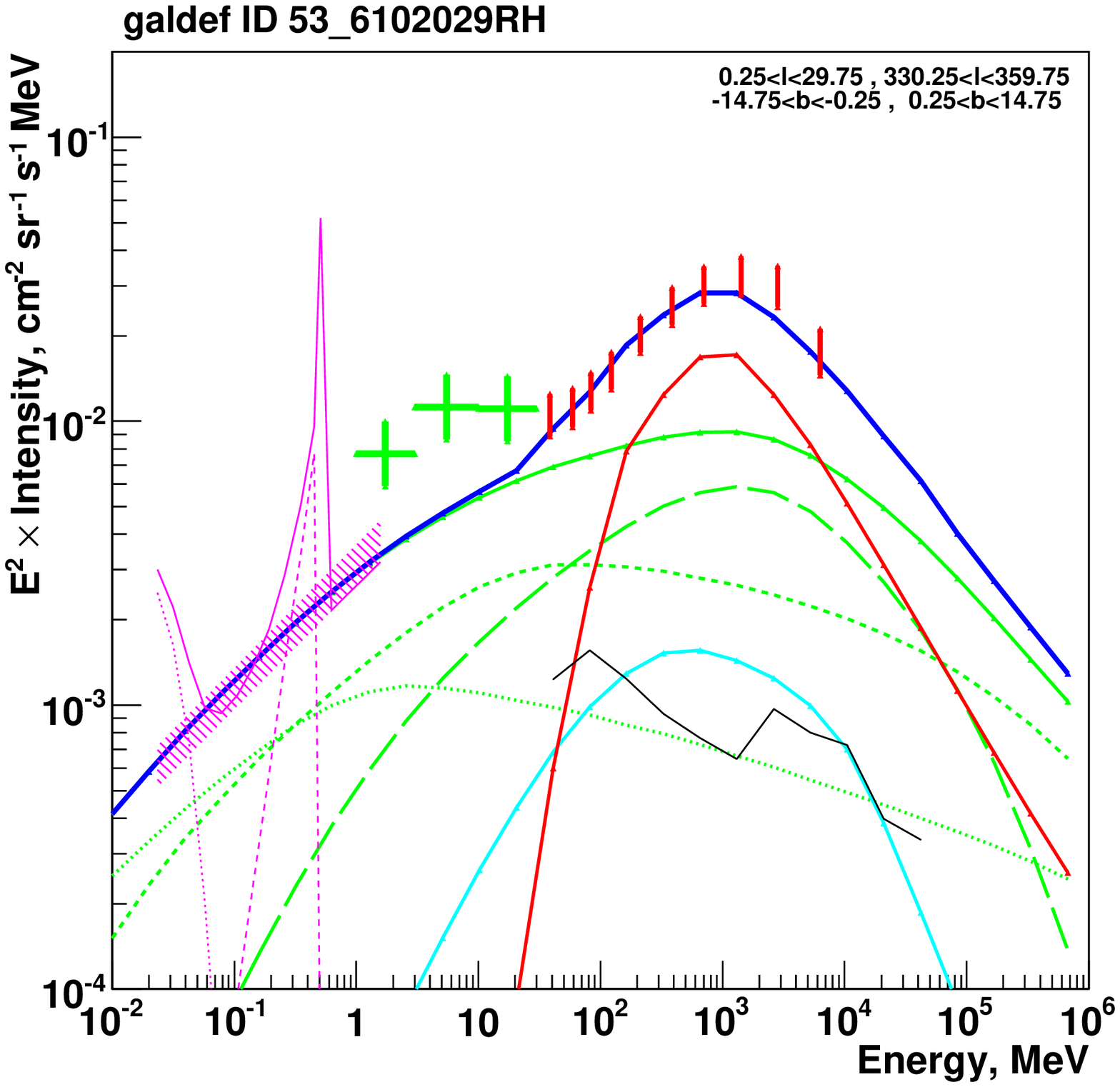}
\includegraphics[width=3.5in]{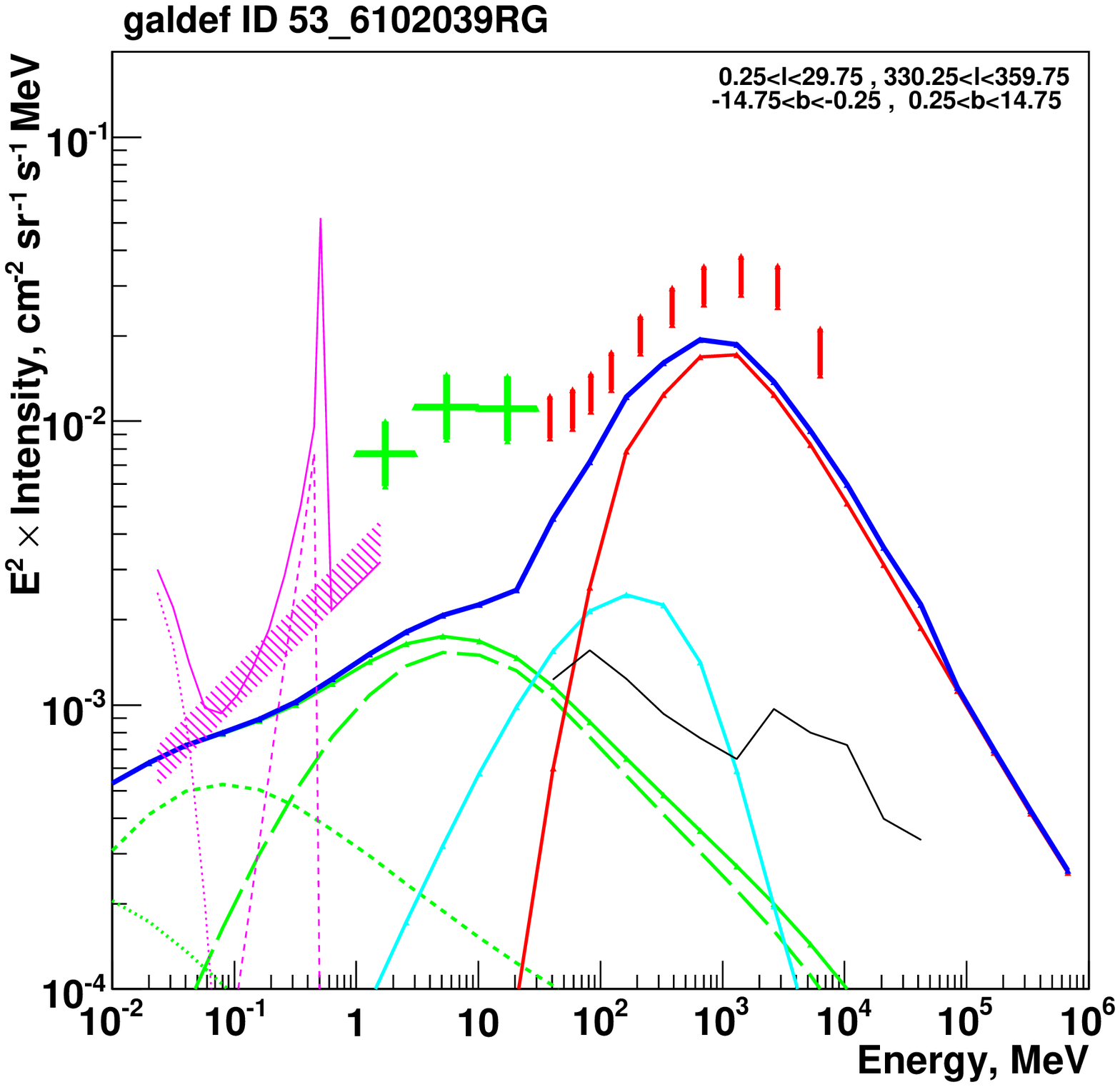}}
\caption{The spectrum of the diffuse emission 
for $330^\circ<l<30^\circ,|b|<15^\circ$ as calculated in the
optimised GALPROP model for the ISRF with maximal metallicity gradient,
{\it Left:} with primary electrons only,
{\it Right:} with secondary electrons and positrons only.
Line-styles: red solid -- $\pi^0$-decay, 
green broken -- IC (optical [long dash], IR [short dash], CMB [dot]), 
green solid -- total IC, 
cyan solid -- bremsstrahlung, black solid -- extragalactic \gray{} background 
\citep{SMR04a}, blue solid -- total. 
Data points: red -- EGRET and green -- COMPTEL, as in \citet{strong05}; 
magenta -- INTEGRAL/SPI (broken lines: components in fit to positronium + 
positron annihilation line + unresolved point sources; shaded region: power-law 
continuum) \citet{bouchet07}.
For the SPI power-law continuum the uncertainty is estimated as described
in the text.
In this and subsequent figures, the identifier (e.g., 53\_6102029RH) 
corresponds
to the GALPROP version and run used; all parameters of the model are contained 
in the ``GALDEF'' parameter file for future reference and are available from 
the GALPROP website at \tt http://galprop.stanford.edu. 
}
\label{fig:diffuse2}
\end{figure*}

We compare our results for the diffuse emission in the inner Galaxy 
with the new SPI data, and COMPTEL and EGRET data.
The spectrum for each instrument is obtained by integrating over deconvolved 
skymaps.
Since the deconvolution of the data is done based on the 
individual instrument response, cross-calibration of the data 
between the individual instruments is not an issue.

Our comparison with the SPI data is with the diffuse emission 
power-law component from \citet{bouchet07}.
We describe the procedure used to obtain the diffuse emission and how 
we estimate the uncertainty on this component.
In the method of \citet{bouchet07}, 
a source catalogue is constructed using an iterative algorithm taking
into account variable source flux contributions using templates for the 
spatial morphologies of the interstellar emission: 
$8^\circ$ degree Gaussian for the positron annihilation emission, 
DIRBE 4.9 $\mu$m and CO maps for the continuum below and above 120 keV, 
respectively.
The normalisation factor for each of these maps is adjusted during the fitting
procedure.
Following this initial step, 
the source fluxes and template information are discarded with only the 
source localisations retained.
The source position information is used in the next step of the analysis where
the region $|l| \leq 100^\circ$ and $|b| \leq 30^\circ$ is divided into cells
with sizes that are chosen to optimise the signal-to-noise ratio per 
cell, while still being sufficiently small to follow the observed spatial 
variations.
A likelihood fit is done using the a-priori source position information to 
obtain the source fluxes and diffuse emission for each `pixel' cell over
the energy ranges 25-50, 50-100, 100-200, 200-600, 600-1800, and 
1800-7800 keV, respectively.
This model independent ``image-based'' method establishes the extent of the 
diffuse emission.
To extract the diffuse spectrum with better signal-to-noise, the 
background templates (DIRBE 4.9 $\mu$m, CO) are also fit for each 
energy range. 
The power-law continuum is based on 
this model-dependent method but there is some error associated 
with the derived emission in this case 
which is not directly estimated in \citet{bouchet07}. 
To estimate the effect on the diffuse emission, we compare the integrated 
latitude profiles obtained using the image-based method and the fit results
for the background template maps given in Fig.5 of \citet{bouchet07}.
We conclude that the intensities could be up to 40\% higher than the
background template ones used by \citet{bouchet07} to construct their spectrum.

The primary electron and secondary positron and electron spectra from 
our propagation calculations are 
shown in Figure~\ref{fig:electrons_positrons}.
For energies $\lesssim 1$ GeV, the combined secondary positron and electron
flux in the ISM actually exceeds the primary electron flux.
This is due to the large ratio of CR nuclei to primary electrons.
The addition of the secondary electrons and positrons increases by 
$\sim 2.5$ the total number capable of producing \gray{s}
via IC scattering relative to the pure primary electrons.

Figure~\ref{fig:diffuse2} shows the individual contributions by 
primary electrons (left) and secondary electrons/positrons (right) to the
diffuse Galactic emission.
For primary electrons, the agreement with the SPI data is excellent while there
is still some deficit when compared with COMPTEL.
For secondary electrons and positrons, 
the spectrum of \gray{s} 
is steeper below $\sim 10$ MeV
compared to the primary electrons, which is a reflection of 
the different source spectra: the primary electron
source spectrum is found from adjusting to the \gray{} spectrum at higher 
energies, while the secondary electron/positron source spectrum
follows from the CR nuclei spectrum in the ISM.
The leptonic component at low energies (IC, bremsstrahlung) is thus 
intrinsically connected with the higher energy 
hadronic component ($\pi^0$-decay).

\begin{figure}
\centerline{
\includegraphics[width=3.5in]{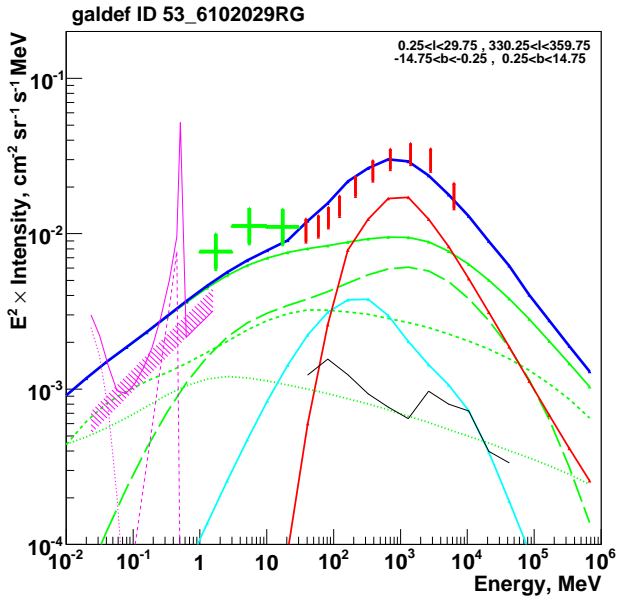}}
\caption{The total spectrum of the diffuse emission as calculated in the
optimised model for $330^\circ<l<30^\circ,|b|<15^\circ$ with the 
maximum metallicity gradient.
Line and data styles as for Fig.~\ref{fig:diffuse2}.
}
\label{fig:diffuse}
\end{figure}

The components of the IC emission (Figure~\ref{fig:diffuse2}) 
show the contributions 
by the ISRF components in different energy ranges.
For primary electrons, the scattering of optical photons is the 
major contribution 
in the energy range $\sim50$~MeV--$100$~GeV, with the infrared the 
major component below $\sim 50$~MeV and above $100$~GeV, and the Cosmic 
Microwave Background (CMB) comparable to the infrared below $\sim$~500 keV.
For the secondary electrons/positrons the scattering of the optical 
component dominates above $\sim 500$~keV, while the infrared is the 
major component for energies below this.
Thus, the primary and secondary populations IC scatter the components of 
the ISRF
to different hard X-ray/\gray{} ranges.
Interestingly, for secondary electrons/positrons the bremsstrahlung
contribution is a factor $\sim 2$ higher for $100$~MeV to $1$~GeV than 
the primary electron case.
This reflects the enhancement of the electrons and positrons 
in the ISM below 1 GeV that 
was discussed in conjunction with Fig.~\ref{fig:electrons_positrons}.

In Figure~\ref{fig:diffuse} we show the diffuse emission calculated  
using the optimised model with an ISRF calculated with a maximal metallicity
gradient.
Inverse Compton scattering is a major component at all energies, with 
$\pi^0$-decay more important between 100 MeV and 10 GeV, 
while the bremsstrahlung
contribution is minor.
The inclusion of secondary electrons and positrons increases the IC emission 
below $\sim 100$ MeV by up to a factor $\sim 2$, an effect that was pointed out 
by \citet{SMR04b}.
Interestingly, the agreement with the COMPTEL data is improved by the
inclusion of the secondary electrons/positrons, although the model still
shows a deficit.
Instead the SPI data are over-predicted, but the spectral slope is still 
consistent with the data below 1~MeV given the estimated uncertainties.
Since the secondary electrons/positrons are a by-product of the same 
processes that produce the $\pi^0$-decay \gray{} emission at higher energies, 
this may indicate that the ratio of CR nuclei to primary electrons we use 
in the ISM is too high.
A possible remedy to recover the model 
fit if the CR nuclei to primary electron ratio is reduced could be that we 
have simply underestimated the optical component of the ISRF.
If the CR nuclei flux is reduced to improve the agreement with
SPI the model emission in the EGRET energy range would also be reduced. 
The emission in the MeV energy range comes from secondary 
electrons/positrons IC scattering the optical component of the ISRF, and 
primary electrons IC scatter the same component to GeV energies.
Increasing the optical ISRF would simultaneously increase the emission in the
MeV and GeV range, recovering the agreement of the model with EGRET data and 
possibly further improving the agreement with COMPTEL.

\begin{figure}
\centerline{
\includegraphics[width=3.5in]{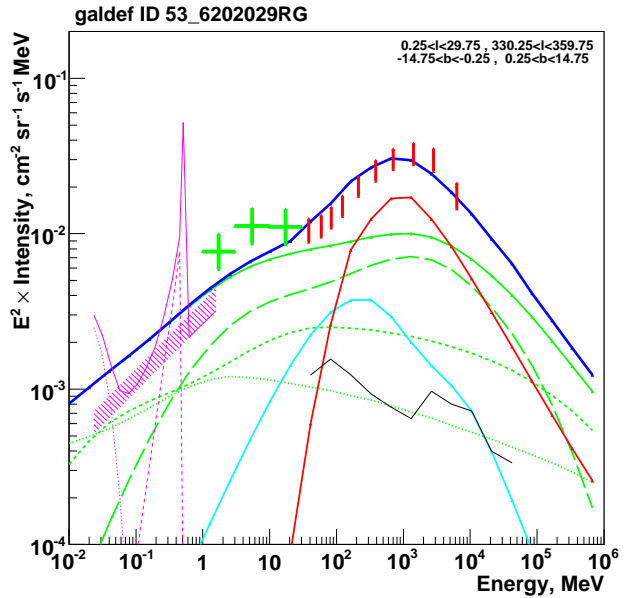}}
\caption{The spectrum of the diffuse emission as calculated in the
optimised model with contribution of secondary electron and positrons and 
ISRF without the metallicity gradient.
Region  $330^\circ<l<30^\circ,|b|<15^\circ$.
Line and data styles as in Fig.~\ref{fig:diffuse2}.}
\label{fig:diffuse_no_metallicity}
\end{figure}

\begin{figure*}
\centerline{
\includegraphics[width=3.5in]{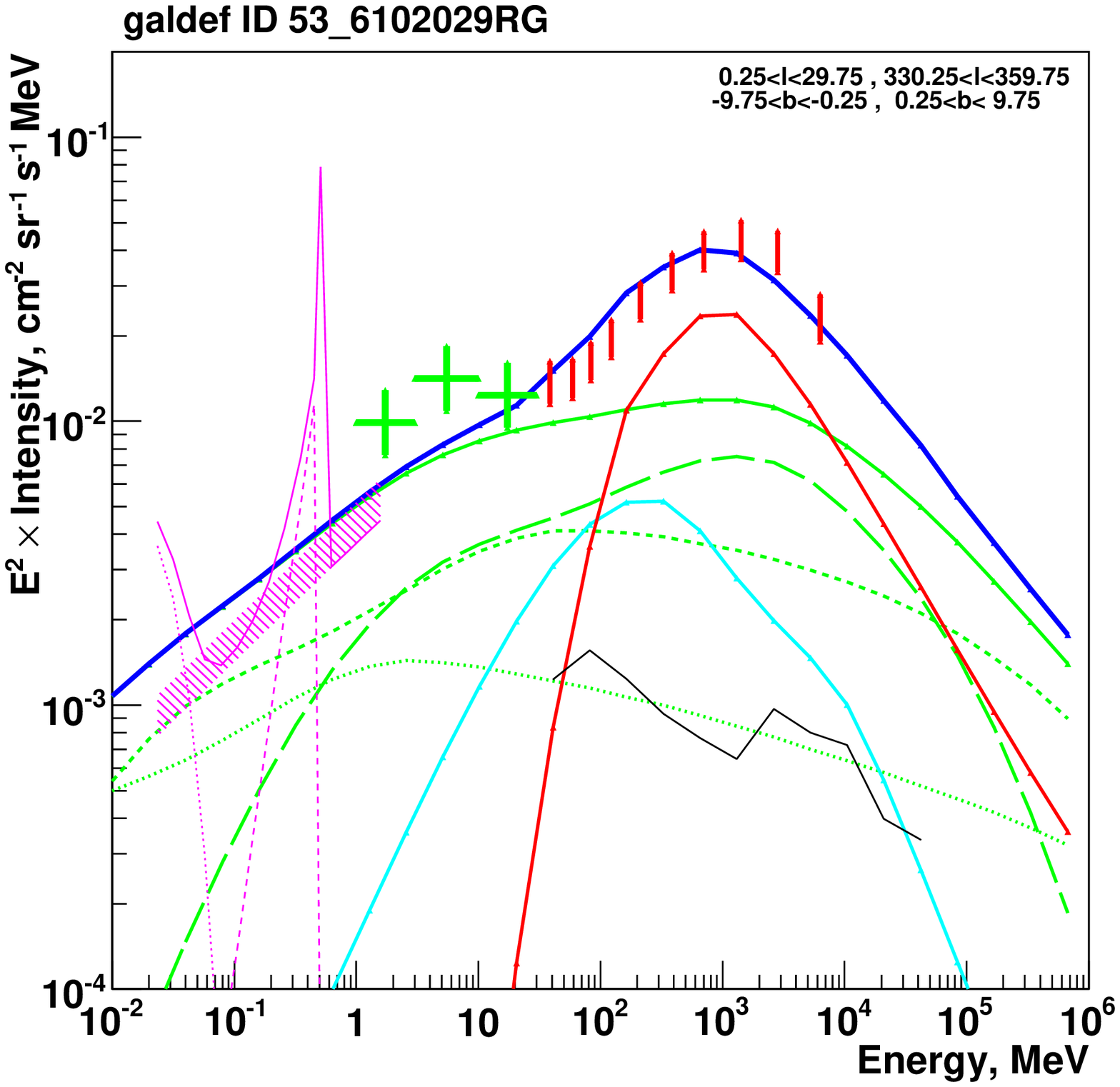}
\includegraphics[width=3.5in]{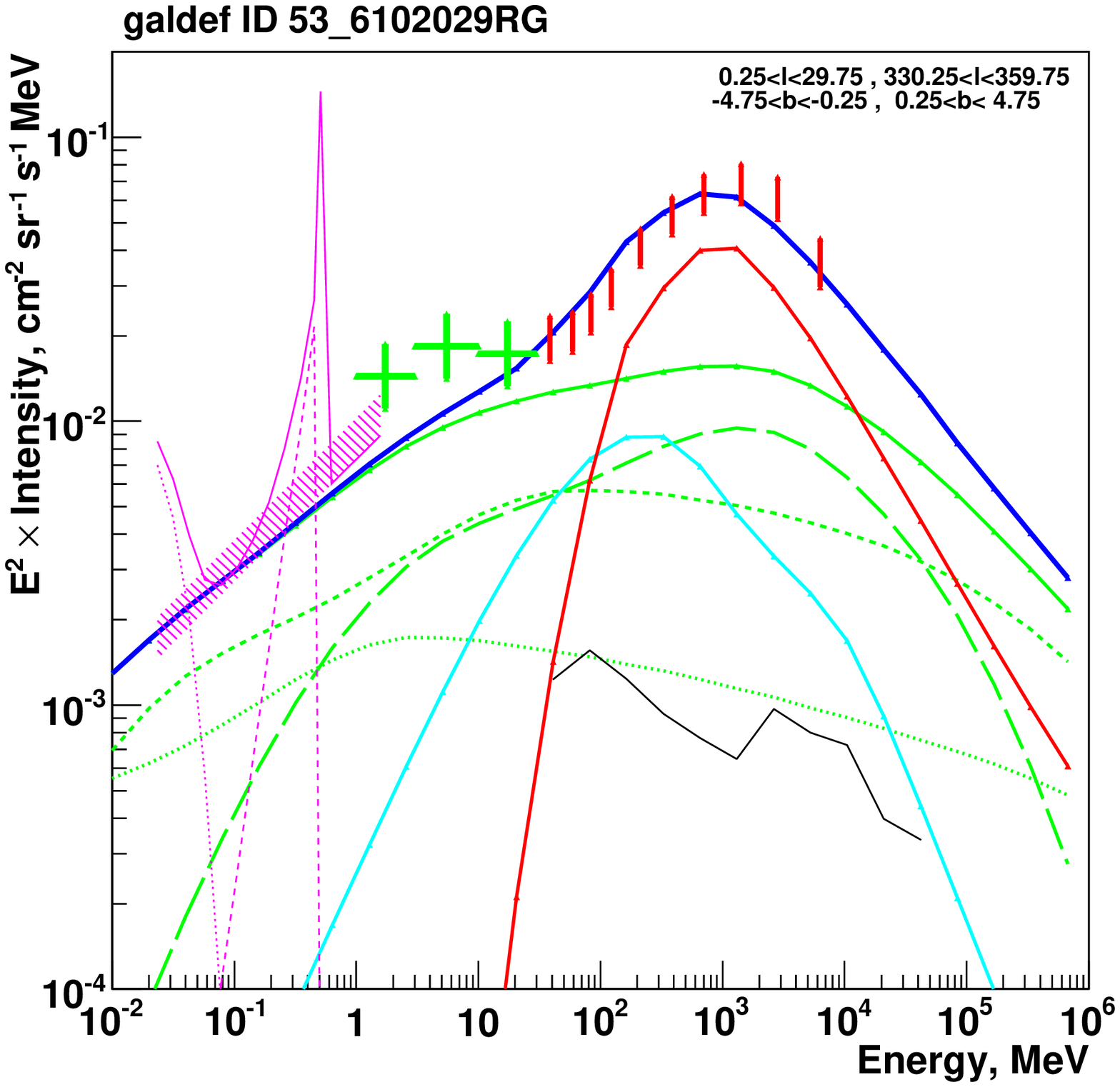}}
\caption{The spectrum of the diffuse emission as calculated in the
optimised model with contribution of secondary electron and positrons 
for different latitude ranges.
{\it Left:} Region $330^\circ<l<30^\circ, |b|<10^\circ$
{\it Right:} Region $330^\circ<l<30^\circ, |b|<5^\circ$.
Line and data styles as in Fig~\ref{fig:diffuse2}.
}
\label{fig:diffuse_other_b}
\end{figure*}

To test the dependence of the sub-MeV emission on the assumed ISRF model, 
we calculate the diffuse
emission for the case of no metallicity gradient.
Figure~\ref{fig:diffuse_no_metallicity} shows the spectrum of the diffuse
emission for this case.
The major change in the ISRF is a $\sim$30\% reduction in the infrared emission 
which is due to the smaller amount of dust in the inner Galaxy when using
this ISRF model.
This results in a drop of $\sim$10--15\% for the 
predicted intensity of the diffuse
emission below $\sim$1 MeV than the maximal gradient case. 
The optical emission increases only slightly, while the CMB emission stays the 
same.
The total spectrum of the diffuse emission does not change significantly 
under this variation of the ISRF showing the robustness of the calculations.

We point out that the reduction in the infrared 
significantly reduces the contribution by secondary
electrons and positrons to the total emission below a few hundred keV.
In this energy range the emission by primary electrons is also reduced, but
by a smaller amount because there is still a contribution by IC scattering 
of the CMB (see Fig.~\ref{fig:diffuse2}).
Since the CMB is known, and if the infrared component of the ISRF is low, 
the emission below a hundred 
keV traces the primary electron spectrum in the ISM.

We have made our principal comparison with the spectrum for $|b|<15^\circ$ 
since this is the nominal range for the spectrum presented in \cite{bouchet07}.
To illustrate the effect of different latitude ranges, we calculate
the diffuse emission for $|b| < 10^\circ$ and $|b| < 5^\circ$ and compare with 
the data for these ranges.
The results for these latitude ranges are shown in 
Fig.~\ref{fig:diffuse_other_b}.
Since the extraction of the diffuse component from the SPI data relies upon
templates whose latitude distribution may not exactly match the true 
distribution of the emission some of the signal may be absorbed 
into the baseline 
(see Fig.~5 of \citet{bouchet07}).
This introduces further uncertainty when comparing reduced latitude ranges 
which is difficult to quantify better than we have already done.
The model emission is qualitatively similar to the 
data for the reduced latitude ranges, which points the way to using the IC 
emission as a template in future analyses of the SPI data.

\section{Discussion and Conclusions}
\label{section:discussion}

The agreement of the single model over the whole energy range from 
SPI data at low energies, to the EGRET data at high energies, is 
remarkable. 
Note that we are using the ``optimised model'' \citep{SMR04b} 
which has higher 
primary CR electron fluxes than observed locally, and also higher
CR nuclei fluxes, as required to fit the \gray{} data above 30 MeV.
With this model we predict too 
much emission below 1 MeV but seem to reproduce the spectral slope. 
Reducing the CR nuclei source spectrum to improve the sub-MeV agreement
is a possibility.
However it cannot be reduced too much since the local CR antiproton fluxes 
must still be reproduced by the 
model\footnote{The BESS-Polar flight of 2004 \citep{Hams2007} 
revealed that the CR antiproton flux is somewhat lower than previous
measurements with lower statistics. 
If these data are confirmed, the CR proton
spectrum will become better constrained.
In turn, it will lead to a somewhat smaller flux of secondary leptons and 
improved agreement with the INTEGRAL data.}.
Increasing the optical component of the ISRF could improve the agreement 
at MeV and GeV energies. 
Variation of the primary electron injection index below a few GeV is 
another possible remedy to the over-production of diffuse emission below 
1~MeV.
This requires a different source spectrum than is presently
used in the optimised model to sufficiently reduce the sub-MeV 
diffuse emission in order to be consistent with 
the SPI spectrum.
If instead we use the ``conventional'' model \citep{SMR04b} 
the situation will not be improved: the sub-MeV emission will be lower, but the 
agreement with the COMPTEL and EGRET data will be substantially worse.

There is still room for a contribution from populations of unresolved 
compact sources, particularly anomalous X-ray pulsars and radio pulsars, 
which may have 
very hard spectra extending to a few hundred keV \citep{Kuiper2006}.
They may be responsible for the apparent peak near $b = 0^\circ$ in the 
\citet{bouchet07} latitude profiles.

The hard X-ray continuum is consistent with the predictions in 
both intensity and spectral index. 
However, the uncertainties in the model are still considerable: the 
distribution of CR sources and gas in the inner Galaxy which affect both 
the primary and secondary electrons/positrons, and the optical and 
infrared part of the ISRF (the CMB is of course known exactly).
In fact, our optimised model overpredicts the SPI data, 
which could simply reflect these uncertainties, but the agreement in the 
spectral shape gives confidence that the 
mechanism is correctly identified.

There is still an excess in the COMPTEL energy range between 1--30 MeV 
which is to be explained. 
The contribution of the positron annihilation in flight
may contribute in this energy range \citep{Beacom06}, but it has to be tested
against the intensity of the 511 keV line and positronium continuum.

From our modelling, we find that the total rate of secondary positron
production by CRs in the whole Galaxy 
is $\sim2\times10^{42}$ s$^{-1}$ in the optimised model. 
The conventional model gives a factor of $\sim$2
less positrons.
These values are $\sim 10$\% of the positron annihilation 
rate $\sim 1.8\times10^{43}$ s$^{-1}$ as derived from INTEGRAL 
observations of the 511 keV line emission \citep{Knodlseder2005}.
The current CR flux of positrons is not sufficient to account for the
observed annihilation rate.
A CR origin for the 511 keV annihilation line can still be reconciled with 
the production rate if CR intensities in the past were higher.

Our work illustrates the intrinsic connection between the diffuse 
Galactic \gray{} emission in different energy ranges.
Inverse Compton emission by CR electrons and positrons on 
starlight and infrared radiation are the most important components
of the hard X-ray and \gray{} emission in the 100 keV to few MeV range.
A considerable proportion of this emission is produced by secondary 
electrons and positrons, the spectrum of which depends on the CR 
nuclei spectrum at energies $\sim$ few GeV and higher.
These CRs also produce $\pi^0$-decay \gray{s} that dominate the emission in 
the GLAST range from 100 MeV to $\sim$10 GeV.
Hence, GLAST observations of the $\pi^0$-decay diffuse emission will also 
constrain in the future the contribution by secondary electrons/positrons 
to the 
diffuse \gray{} emission in the SPI energy range.
With the secondary electrons/positrons fixed, SPI observations probe the 
IC emission of primary CR electrons with energies $\lesssim 10$ GeV scattering
the infrared component of the ISRF and the CMB.
This will provide information on the low energy spectrum of primary CR 
electrons and the infrared component of the ISRF. 
In turn, since most of the diffuse \gray{} emission between 
$\sim$10 GeV -- 10 TeV is produced via IC scattering of primary electrons 
on the same starlight and infrared photons,   
this provides a connection to observations of diffuse emission at TeV 
energies by HESS \citep{aharonian2006} and MILAGRO \citep{abdo2007}.

\acknowledgments
T.\ A.\ P.\ acknowledges partial support from the US Department of Energy.
I.\ V.\ M.\ acknowledges partial support from NASA
Astronomy and Physics Research and Analysis Program (APRA) grant.

\clearpage

\clearpage

\end{document}